\documentstyle[12pt,epsf]{article}

\catcode`\@=11
\long\def\@makefntext#1{
\protect\noindent \hbox to 3.2pt {\hskip-.9pt  
$^{{\ninerm\@thefnmark}}$\hfil}#1\hfill}		

\def\@makefnmark{\hbox to 0pt{$^{\@thefnmark}$\hss}}  
	
\def\ps@myheadings{\let\@mkboth\@gobbletwo
\def\@oddhead{\hbox{}
\rightmark\hfil\ninerm\thepage}   
\def\@oddfoot{}\def\@evenhead{\ninerm\thepage\hfil
\leftmark\hbox{}}\def\@evenfoot{}
\def\sectionmark##1{}\def\subsectionmark##1{}}

\renewcommand{\thefootnote}{\fnsymbol{footnote}}

\newcounter{sectionc}\newcounter{subsectionc}\newcounter{subsubsectionc}
\renewcommand{\section}[1] {\vspace*{0.6cm}\addtocounter{sectionc}{1} 
\setcounter{subsectionc}{0}\setcounter{subsubsectionc}{0}\noindent 
	{\normalsize\bf\thesectionc. #1}\par\vspace*{0.4cm}}
\renewcommand{\subsection}[1] {\vspace*{0.6cm}\addtocounter{subsectionc}{1} 
	\setcounter{subsubsectionc}{0}\noindent 
	{\normalsize\it\thesectionc.\thesubsectionc. #1}\par\vspace*{0.4cm}}
\renewcommand{\subsubsection}[1]
{\vspace*{0.6cm}\addtocounter{subsubsectionc}{1}
	\noindent {\normalsize\rm\thesectionc.\thesubsectionc.\thesubsubsectionc. 
	#1}\par\vspace*{0.4cm}}

\newcounter{appendixc}
\newcounter{subappendixc}[appendixc]
\newcounter{subsubappendixc}[subappendixc]

\renewcommand{\appendix}[1] {\vspace*{0.6cm}
        \refstepcounter{appendixc}
        \setcounter{figure}{0}
        \setcounter{table}{0}
        \setcounter{equation}{0}
        \renewcommand{\thefigure}{\Alph{appendixc}.\arabic{figure}}
        \renewcommand{\thetable}{\Alph{appendixc}.\arabic{table}}
        \renewcommand{\theappendixc}{\Alph{appendixc}}
        \renewcommand{\theequation}{\Alph{appendixc}.\arabic{equation}}
        \noindent{\bf Appendix \theappendixc #1}\par\vspace*{0.4cm}}

\def\abstracts#1{{
	\centering{\begin{minipage}{12.2truecm}\footnotesize\baselineskip=12pt\noindent
	\centerline{\footnotesize ABSTRACT}\vspace*{0.3cm}
	\parindent=0pt #1
	\end{minipage}}\par}} 

\renewenvironment{thebibliography}[1]
	{\begin{list}{\arabic{enumi}.}
	{\usecounter{enumi}\setlength{\parsep}{0pt}
\setlength{\leftmargin 1.25cm}{\rightmargin 0pt}
	 \setlength{\itemsep}{0pt} \settowidth
	{\labelwidth}{#1.}\sloppy}}{\end{list}}

\newcounter{itemlistc}
\newcounter{romanlistc}
\newcounter{alphlistc}
\newcounter{arabiclistc}

\newcommand{\fcaption}[1]{
        \refstepcounter{figure}
        \setbox\@tempboxa = \hbox{\footnotesize Fig.~\thefigure. #1}
        \ifdim \wd\@tempboxa > 6in
           {\begin{center}
        \parbox{6in}{\footnotesize\baselineskip=12pt Fig.~\thefigure. #1}
            \end{center}}
        \else
             {\begin{center}
             {\footnotesize Fig.~\thefigure. #1}
              \end{center}}
        \fi}

\newcommand{\tcaption}[1]{
        \refstepcounter{table}
        \setbox\@tempboxa = \hbox{\footnotesize Table~\thetable. #1}
        \ifdim \wd\@tempboxa > 6in
           {\begin{center}
        \parbox{6in}{\footnotesize\baselineskip=12pt Table~\thetable. #1}
            \end{center}}
        \else
             {\begin{center}
             {\footnotesize Table~\thetable. #1}
              \end{center}}
        \fi}

\def\@citex[#1]#2{\if@filesw\immediate\write\@auxout
	{\string\citation{#2}}\fi
\def\@citea{}\@cite{\@for\@citeb:=#2\do
	{\@citea\def\@citea{,}\@ifundefined
	{b@\@citeb}{{\bf ?}\@warning
	{Citation `\@citeb' on page \thepage \space undefined}}
	{\csname b@\@citeb\endcsname}}}{#1}}

\newif\if@cghi
\def\cite{\@cghitrue\@ifnextchar [{\@tempswatrue
	\@citex}{\@tempswafalse\@citex[]}}
\def\citelow{\@cghifalse\@ifnextchar [{\@tempswatrue
	\@citex}{\@tempswafalse\@citex[]}}
\def\@cite#1#2{{$\null^{#1}$\if@tempswa\typeout
	{IJCGA warning: optional citation argument 
	ignored: `#2'} \fi}}

 1
 1
 1

\font\ninerm=cmr9

\begin{document}

\centerline{\normalsize\bf RANDOMLY CHARGED POLYMERS: EXCESS CHARGE}
\baselineskip=22pt
\centerline{\normalsize\bf DEPENDENCE OF SPATIAL CONFIGURATIONS}
\baselineskip=16pt

\vspace*{0.6cm}
\centerline{\footnotesize YACOV KANTOR}
\baselineskip=13pt
\centerline{\footnotesize\it School of Physics \& Astronomy, Tel Aviv University}
\baselineskip=12pt
\centerline{\footnotesize\it 69978 Tel Aviv, Israel}
\centerline{\footnotesize E-mail: kantor@orion.tau.ac.il}
\vspace*{0.3cm}
\centerline{\footnotesize and}
\vspace*{0.3cm}
\centerline{\footnotesize MEHRAN KARDAR}
\baselineskip=13pt
\centerline{\footnotesize\it Department of Physics, Massachusetts 
Institute of Technology}
\baselineskip=12pt
\centerline{\footnotesize\it Cambridge, MA 02139, U.S.A.}
\centerline{\footnotesize E-mail: kardar@mit.edu}

\vspace*{0.9cm}
\abstracts{
Spatial configurations of randomly charged polymers, known as 
poly\-ampho\-lytes (PAs), are very sensitive to the overall excess 
charge $Q$. Analytical arguments, supported by Monte Carlo 
simulations\cite{KKL,KKMC} and exact enumeration studies,\cite{KKenum}
lead to the following picture: For $Q<Q_c\approx q_0\sqrt{N}$ ($q_0$ is
the elementary charge, $N$ is the number of monomers in the polymer),
the radius of gyration $R_g$ of the polymer decreases with decreasing 
temperature $T$ and the polymer becomes compact, while for $Q>Q_c$ the
polymer stretches with decreasing $T$. At low $T$, the dense states are
described by Debye--H\"uckel theory, while the expanded states resemble
a {\em necklace} of globules connected by strings. At such temperatures, the
transition between the dense and the expanded states with increasing 
$Q$, is reminiscent of the breakup of a charged drop.
}
 
\normalsize\baselineskip=15pt
\setcounter{footnote}{0}
\renewcommand{\thefootnote}{\alph{footnote}}
\section{Introduction}

Statistical mechanics of randomly charged polymers, called 
{\em polyampholytes} (PAs), is a challenging subject because it embodies an
interesting combination of long range interactions and randomness.
While the physics of homogeneous polymers has a reasonably firm
basis,\cite{rPolGen} considerably less is known about heteropolymers,
although the latter present an extremely rich problem of
biological significance.\cite{multi}  In this presentation we
review some properties of PAs.  

The spatial extent of a polymer is characterized by the critical exponent
$\nu$, which relates its radius of gyration (r.m.s. size) $R_g$ to the
number of monomers $N$, by the power law $R_g\propto N^\nu$. The polymer
will be called ``compact'' if $\nu=1/d$, where $d$ is the dimension of
the embedding space, and ``stretched'' if $\nu=1$.
The simplest model of PAs is as a flexible chain of $N$ monomers, 
each of which
has a  charge $\pm q_0$ selected from a well defined 
ensemble of quenches. The polymer has a characteristic microscopic
length scale $a$, such as the range of excluded volume interactions, 
or the nearest neighbor distance along the chain. The monomers of the
PA interact both via (short range) excluded volume interactions
and long range (unscreened) Coulomb interactions.
In the simplest ensemble of quenches, each monomer takes a 
charge $q_i=\pm q_0$ independently of all the others; i.e. 
$\overline{q_iq_j}=\delta_{ij}q_0^2$, where the overline 
indicates averaging over quenches.
While the average excess charge $Q\equiv\sum_iq_i$ of such PAs is 
zero, a ``typical''
sequence has Q of about $\pm Q_c$,
with $Q_c\equiv q_0N^{1/2}$. This statement, as well 
as the definition of $Q_c$, is unrelated to the embedding 
dimension $d$. However, the importance of charge fluctuations 
(both for the overall polymer, or for large segments of it) 
does depend on the space dimension.
The electrostatic energy of the excess charge,
spread over the characteristic size of an ideal polymer
($R_g\propto N^{1/2}$), grows as $Q^2/R_g^{d-2}\sim N^{(4-d)/2}$.
This simple dimensional argument shows that for $d>4$ weak 
electrostatic interactions are irrelevant. (The excluded volume 
effects are also irrelevant in $d>4$.)
Thus, in $d>4$ and at high temperatures, the PA behaves as an ideal polymer 
with an entropy--dominated free energy of the order of $-NT$. However, 
on lowering temperature it collapses  into a dense state, 
taking advantage  of a condensation energy of the order of 
$-Nq_0^2/a^{d-2}$. 

We are primarily interested in the behavior of PAs in $d=3$.
We present both numerical and analytical evidence that the behavior
of PA in $d<4$ is extremely sensitive to the value of $Q$. In particular,
we show that for $Q<Q_c$ the PA assumes compact configurations, while for
$Q>Q_c$ it is in an expanded state. We suggest a ``necklace''
model which qualitatively describes the latter state.

\section{Properties of Polyampholytes}
\subsection{Neutral Heteropolymers: Short--Range {\em vs.} Long--Range
Interactions}

\begin{figure}
\vspace*{13pt}
\centerline{\epsfysize=2.5truein\epsffile{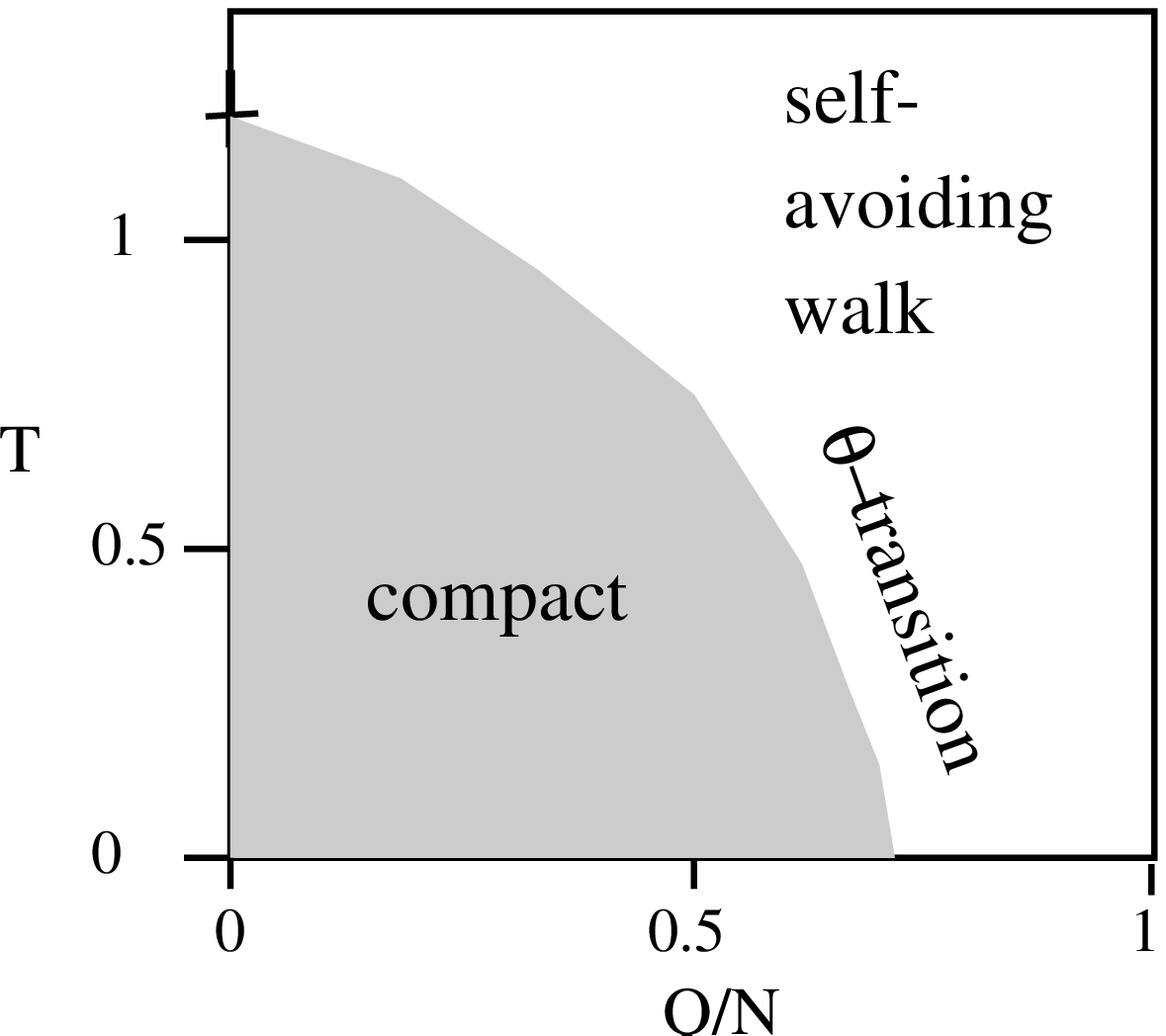}
\hskip 1cm\epsfysize=2.5truein\epsffile{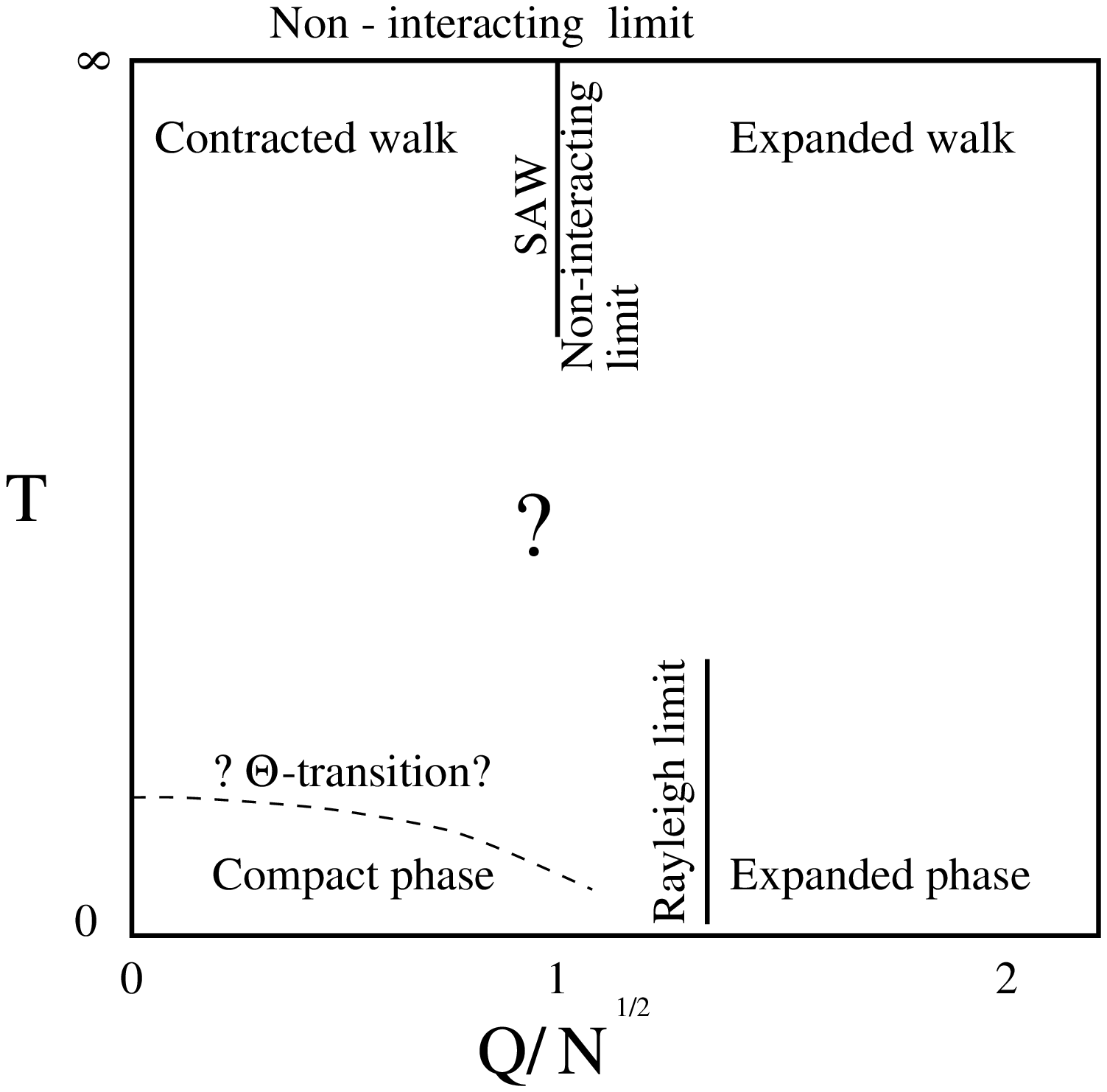}}
\centerline{\hskip 1.4cm (a) \hskip 6.3truecm (b) }
\fcaption{Phase diagrams for random polymers with (a) short--range, 
and (b) Coulomb interactions, in $d=3$. Note the different scaling of the
excess charge $Q$ in the two figures.}
\label{fig:diagram}
\end{figure}

It is interesting to compare and contrast the behaviors of heteropolymers
with short  and long range interactions. Consider a polymer represented
by a self--avoiding chain carrying a quenched sequence of random charges
which interact only at {\em short} distances, with similar charges repelling
each other and the opposite charges attracting.\cite{rsrim} 
In $d=3$, at high
temperatures $T$, the presence of such short range interactions does not
influence the behavior of the polymer and it behaves as a self--avoiding
walk (SAW).  If  $Q$ is small, i.e. the positive and negative charges
are almost balanced, upon decrease of temperature, charges 
of opposite sign are more likely to be in the vicinity of each other.
This introduces an effective attraction which increases in strength 
with decreasing temperature, and at a certain temperature the polymer
collapses into a globule, as in a regular $\theta$--transition.
When the imbalance $Q$ between the two types of charges increases,
the transition temperature decreases, as demonstrated in 
Fig.~\ref{fig:diagram}a, and for sufficiently large imbalance the transition
disappears altogether.\cite{rsrim} Since the behavior of the system varies very
gradually with increasing $Q/N$, it does not really matter whether the
ensemble of chains consists only of quenches with exactly vanishing $Q$, 
or is obtained by independently choosing each 
$q_i=\pm q_0$ with equal probability. In the latter case $Q$ may deviate
from zero by an amount of order $q_0N^{1/2}$; such fluctuations have a 
negligible contribution to $Q/N$ for large $N$.

The long range Coulomb interactions in random PAs are always 
relevant at $d<4$. An approximate treatment using a Debye--H\"uckel type
theory\cite{rHJ} leads to the conclusion that a PA minimizes its
free energy by assuming a configuration in which positive charges
are predominantly surrounded by negative charges, causing screening
as in a regular electrolyte. A detailed calculation suggests\cite{rHJ,KKL}
the following picture: At a finite $T$, the PA can be divided into
segments containing $n$ monomers each, where $n$ defines the length
scale at which electrostatic interactions become of order $T$, i.e.
$Q^2(n)/R_g(n)\sim q_0^2n/(an^{\nu_0})=T$ in  $d=3$, where $\nu_0\approx 0.59$.
The overall configuration is then composed of $n$--mer blobs: Inside
the blob the PA behaves as an uncharged polymer, while the blobs form
a compact structure resembling an electrolyte. The resulting PA
is compact (i.e., has $R_g\sim N^{1/3}$) at all temperatures,
with a $T$--dependent density. On the other hand, scaling arguments\cite{rKK}
assume that the $N$--dependence of the $R_g$ of a random PA is
determined from the relation $Q^2/R_g\sim q_0^2N/R_g\approx T$,
leading to the conclusion that $R_g\sim N$ in $d=3$.
This apparent contradiction is resolved by noting\cite{KKL} that the
spatial configurations of PAs are extremely sensitive to the overall
charge $Q$. In particular, it can be shown\cite{KKL,KKMC}
 that there is a critical
charge $Q_c=q_0N$, such that for $Q<Q_c$ the PAs are compact, while for
$Q>Q_c$ they are stretched. The diagram in 
Fig.~\ref{fig:diagram}b indicates such separation between contracted
and expanded regimes of the PA. In the following subsections we shall
present different arguments and numerical results supporting this conclusion.

\subsection{Polyampholytes with Excess Charge: High Temperature Arguments}

For $d<4$, electrostatic interactions are relevant and the 
high temperature phase is no longer a regular self--avoiding 
walk. At high $T$, the behavior of the polymer can be 
explored perturbatively. For the ensemble of uncorrelated
charges ($\overline{q_iq_j}=0$), the lowest order 
($1/T$) correction to the quench--averaged $R_g^2$ vanishes.\cite{KKL}
However, restricting the ensemble to 
yield fixed $Q$, slightly modifies the quench--averaged
charge--charge correlations. In
particular, the two--point correlation function becomes
$\overline{q_iq_j}=(Q^2-Q_c^2)/N^2$ for $i\ne j$.
This small (order of $1/N$) correction to the correlation 
function causes a significant change in $R_g^2$ due to the
long range nature of the Coulomb interaction.
A $1/T$--expansion\cite{KKL} has a first order
term  proportional to $Q-Q_c$. Thus the size of a PA tends to 
decrease upon lowering temperature if $Q$ is less than
$Q_c$, and increases otherwise.

Confirming the $1/T$--expansion, a strong $Q$--dependence 
of $R_g$ has been seen in Monte Carlo simulations:\cite{KKMC} It has been 
shown\cite{KKL} that for $Q<Q_c$ the PA contracts with decreasing $T$,
while for $Q>Q_c$ it expands. Fig.~\ref{fig:config} depicts 
low temperature equilibrium configurations of two PAs, one of them has $Q$ slightly
larger than $Q_c$ and is strongly stretched, while the other has
$Q=0$ and forms a globule. The two behaviors
are separated by the vertical line at the
top of the phase diagram in Fig.~\ref{fig:diagram}b. It should be
noted that the arguments leading to location of this line are
independent of $d$, as long as $d<4$.

\begin{figure}
\vspace*{13pt}
\centerline{\epsfysize=2.5truein\epsffile{Fig4d.epsi}
\hskip 4cm\epsfysize=2.5truein \epsffile{Fig4a.epsi}}
\centerline{\hskip 1.3cm (a) \hskip 8.5truecm (b) }
\fcaption{Low $T$ configuration of 64--monomer PAs with (a) $Q=12$,
and (b) $Q=0$. Dark and bright shades indicate opposite charges.}
\label{fig:config}
\end{figure}

\subsection{Polyampholytes with Excess Charge: Low Temperature Arguments}

\begin{figure}
\vspace*{13pt}
\epsfysize=3.0truein\centerline{\epsffile{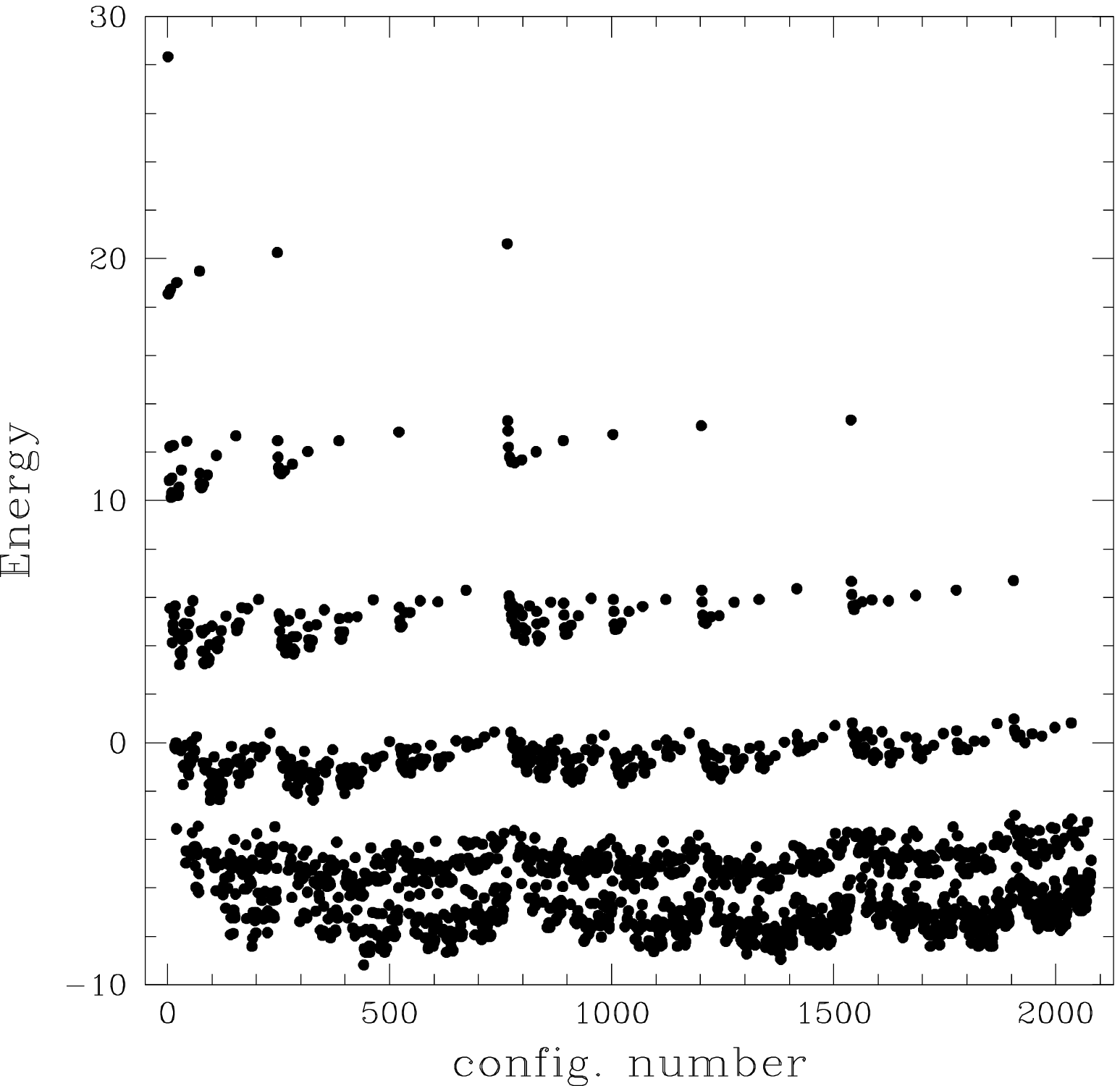}}
\fcaption{Ground state energies of (arbitrarily numbered) quenches.}
\label{fig:spectrum}
\end{figure}

Monte Carlo results indicate that  PAs with $Q=0$ form
dense globules at low temperatures. We can use this observation 
as a starting point 
for the investigation of the dependence of low--$T$ shapes
on $Q$ in $d=3$.  The energy (or rather the 
quench--averaged free energy) of the PA is phenomenologically 
related to its shape by
\begin{equation}\label{edefenerg}
E=-\epsilon_cN+\gamma S + {b Q^2\over R}.
\end{equation}
The first term is a condensation energy proportional to the 
volume (assumed compact) and $\epsilon_c\sim q_0^2/a$, 
the second term is proportional to the
surface area $S$ (with a surface tension $\gamma\sim q_0^2/a^3$), 
while the 
third term represents the long--range part of the electrostatic 
energy due to an excess charge $Q$ ($b$ is a dimensionless 
constant of order unity). The correctness of separating the energy
into the 3 terms appearing in Eq.~\ref{edefenerg} is not self--evident
in {\em random} systems. A justification for the assumptions
implicit in this equation is provided by an exact enumeration
study of the ground states of PAs defined on a lattice:\cite{KKenum}
Fig.~\ref{fig:spectrum} depicts the ground state energies of {\it all} 2080 
possible quenches for 12--step (13--atom) chains. 
(The horizontal axis represents an arbitrary numbering of the quenches.)
The energies are clearly separated into 7 bands, corresponding to 
excess charges of  $Q=1$, 3, 5, $\cdots$, 13. (There is only
one quench with $Q=13$.) While each band has a finite width, we 
see that the energy of a PA can be determined rather accurately
by only specifying its net charge $Q$, while the additional details 
of the quench appear only in the width of the band.

The optimal shape is obtained by 
minimizing the overall energy in Eq.~\ref{edefenerg}. 
The first term is the same for
all compact shapes, while the competition between the surface
and electrostatic energies is controlled by the dimensionless
parameter
\begin{equation}
\alpha\equiv {Q^2\over 16\pi R^3\gamma}=
{Q^2\over 12V\gamma}\equiv {Q^2\over Q_R^2}\ .
\end{equation}
Here, $R$ and $V$ are the radius and volume of a spherical drop
of $N$ particles, and we have defined the {\it Rayleigh charge} 
$Q_R$. We note that in the case of PAs\cite{KKMC,KKenum} 
$Q_R\approx q_0N^{1/2}=Q_c$.
The behavior of a system described by Eq.~\ref{edefenerg}
has been analyzed in the past in the contexts of charged conducting
drops, as well as  liquid drop models of atomic 
nuclei. If one restricts the possible shapes of a globule
to spheroids, then one finds that for sufficiently large values of 
$\alpha$ (of order one or more), the energy of the system can be
decreased by distorting the drop into a prolate spheroidal shape.
Both MC simulations\cite{KKMC} and experimental results\cite{rExT} 
indicate that for $Q<Q_R$, the $R_g$ of PA at low temperature 
is almost independent of $Q$, while
it increases extremely fast as a function of $Q$ for larger charges.

The transition between compact and expanded states with increasing
$Q$ is represented by the vertical line at the bottom of the diagram
in Fig.~\ref{fig:diagram}b. The presence of this line signifies the 
instability of the spherical shape, but provides no indication of what
the stable state of the PA looks like. This question is taken up
in the following section. It should be stressed, that although both
high and low temperature arguments seem to provide a consistent picture
of transition between compact and extended states, the equality between
$Q_c$ and $Q_R$ depends on the space dimension. For $d<3$ ($d>3$)
$Q_R$ increases with $N$ slower (faster) than
$Q_c$, and therefore the high--$T$ criterion will not coincide
with the low--$T$ limit.

\section{The Necklace Model}

\subsection{Uniform Charges}

Although the ellipsoidal globule may have a lower energy than
the spherical one, the former is {\em not} the minimum energy
configuration of a charged drop: A uniformly charged
drop minimizes its energy by {\em splitting} into two equal, 
infinitely removed,  droplets for $\alpha>0.3$, and
the number of such droplets increases with $\alpha$. Obviously,
the PA must maintain its connectivity. We can 
constrain the overall shape to  remain singly 
connected by linking the droplets via narrow tubes of total  
length $L$ and diameter $a$. As long as $L a^2\ll R^3$, most 
of the charge  remains in the spheres. The total 
electrostatic energy is proportional to $Q^2/L$,
while the surface energy cost grows as $\gamma a L$; equating 
the two gives $L\propto Q$. For large $Q$, the PA will look
like a {\em necklace} of globules connected by narrow strands.
The configuration depicted in Fig.~\ref{fig:config}a indeed bears
some resemblance to such a necklace. 

It should be noted that $R_g$ in this simplified discussion is 
proportional to the typical charge $Q$, and therefore proportional
to $N^{1/2}$.
This serves as an indication that the quench--averaged 
configurations are not compact. The picture has one important 
shortcoming for PAs:
it assumes that the excess charge is uniformly distributed
along the chain and disregards the strong charge fluctuations.
It thus applies to such problems as an alternating sequence of
charges\cite{victor} to which some additional charge has been
added, or to a weakly charged homogeneous polymer
in which the attraction (and, therefore, the condensation
energy) is provided by a different mechanism, such as short range
attractions.

\subsection{Effects of Randomness}

We now illustrate the difficulties caused by randomness for 
the case of an unrestricted PA. Since $\overline{Q^2}=q_0^2N$, 
where the overline denotes an average over the ensemble of 
all quenches, we have $\overline{\alpha}=1$. A uniformly charged 
drop is unstable 
to splitting already for $\alpha\approx0.3$, and thus a 
typical random PA is expected to form several globules 
connected by narrow tubes. Now consider splitting  sequences 
of $N$ monomers with total charge constrained by a particular 
$\alpha$ into two equal subchains of charges $Q_1$ and $Q_2$. 
It is easy to show that each segment has
$\overline{\alpha_{\rm subchain}}=(1+\alpha)/2$, while the mean  
product of the charges is $\overline{Q_1Q_2}=q_0^2N(\alpha-1)/4$. 
The subchains have, on average, values of $\alpha$ close to 
unity. Also, for $\alpha=1$, the  average value of the product
of charges vanishes. We thus have the paradoxical situation in 
which most spherical shapes are unstable, while there is on 
average no energetic gain in splitting the sphere into two 
equal parts. 

Therefore, charge inhomogeneities drastically modify the necklace  
picture. The resulting PA is probably  still composed of rather 
compact globules connected by a (not necessarily linear) 
network of tubes. The globules are selected preferentially from 
segments of the chain that are approximately neutral (or at 
least below the instability threshold), while the tubes are 
from subsequences with larger than average excess charge. 
It can be shown\cite{KE} that the probability
of finding a very large neutral segment in a random sequence
of charges is large, i.e. it is possible to build a configuration
consisting of one very large neutral globule with highly
charged ``tails'' sticking out of it. 

While the necklace model provides a convenient starting picture
for the behavior of PAs, it does not encompass the entire complexity
of the problem. The probability distribution of the $R_g$ of
the ground states of PAs obtained from exact enumeration 
study\cite{KKenum} has a large peak for small $R_g$,
indicating that most of the configurations are compact, as well as
a slowly decaying tail at large $R_g$. This tail
represents the expanded configurations, and determines the behavior of
the quench--averaged $R_g$. Indeed, for $N\le13$ we found\cite{KKenum}
that 80\% of all quenches can be classified as compact, i.e. their
$R_g$ increases as $N^{1/3}$, while the remaining 20\% have quench--averaged
$R_g$s increasing with $N$ at least as fast as that of a SAW.

We can summarize our current knowledge of the behavior of random PAs
as follows:
The exact enumeration and Monte Carlo studies confirm that random
PAs have (on the average) expanded spatial conformations, although
further work is needed to verify whether $\nu=1$. The necklace
model provides a useful qualitative view of the ground state configurations.
Further studies are needed to put the model on a more quantitative basis.

\section{Acknowledgements}
This work was supported by the US--Israel BSF grant 
No. 92--00026, and by the NSF through grant No. DMR--94--00334 at MIT's 
CMSE.

\medskip


\section{References}

\begin{thebibliography}{9}


\bibitem{rPolGen}{P.G. de Gennes, {\it Scaling Concepts in Polymer Physics},
Cornell Univ. Press, Ithaca (1979).}
\bibitem{multi}{D.L. Stein, Proc. Natl. Acad. Sci. USA {\bf 82},
3670 (1985);
J.D. Bryngelson and P.G. Wolynes, Proc. Natl. Acad. Sci. USA
{\bf 84}, 7524 (1987); H.S. Chan and K.A. Dill, Physics Today {\bf 46}(2), 
24 (1993); T. Garel and H. Orland, Europhys. Lett. {\bf 6}, 307  
(1988);
E.I. Shakhnovich and A.M. Gutin, Europhys. Lett. {\bf 8}, 327 (1989);
M. Karplus and E.I. Shakhnovich, in {\it Protein Folding}, ed. by 
T.E. Creighton, ch.4,
p. 127, (Freeman \& Co., New York, 1992).}
\bibitem{rsrim}Y. Kantor and M. Kardar, Europhys.
Lett. {\bf 28}, 169 (1994).
\bibitem{rHJ}P.G. Higgs and J.-F. Joanny, J. Chem. Phys. {\bf 94},  
1543 (1991); J. Wittmer, A. Johner and J.F. Joanny, Europhys.
Lett. {\bf 24}, 263 (1993).
\bibitem{KKL}{Y. Kantor, H. Li, and M. Kardar, Phys. Rev. Lett.  
{\bf 69}, 61 (1992); Y. Kantor, M. Kardar, and H. Li, Phys. Rev. {\bf  
E49}, 1383 (1994).}
\bibitem{rKK}{Y. Kantor and M. Kardar, Europhys. Lett. {\bf 14}, 421  
(1991).}
\bibitem{KKMC}Y. Kantor and M. Kardar,  Europhys.  
Lett. {\bf 27}, 643 (1994); Y. Kantor and M. Kardar, Phys. Rev.  
{\bf E51}, 1299 (1995).
\bibitem{KKenum}Y. Kantor and M. Kardar, Phys. Rev.  
{\bf E52}, in press (1995).
\bibitem{rExT}X.-H. Yu, A. Tanaka, K. Tanaka, and T. Tanaka, 
J. Chem. Phys. {\bf 97}, 7805 (1992);
Yu X.-H., Ph. D. thesis, MIT (1993).
\bibitem{victor}J.M. Victor and J.B. Imbert, Europhys. Lett.
{\bf 24}, 189 (1993).
\bibitem{KE}Y. Kantor and D. Erta\c s, J. Phys. {\bf A27}, L907 (1994);
D. Erta\c s and Y. Kantor, Phys. Rev. {\bf E}, submitted (1995).

\end{thebibliography}
\end{document}